# Probabilistic forecasting for geosteering in fluvial successions using a generative adversarial network

Sergey Alyaev (saly@norceresearch.no NORCE Norwegian Research Centre, P.O.B 22 Nygårdstangen NO-5838 Bergen, Norway);
Jan Tveranger (NORCE, P.O.B 22 Nygårdstangen NO-5838 Bergen, Norway);
Kristian Fossum (NORCE, P.O.B 22 Nygårdstangen NO-5838 Bergen, Norway);
Ahmed H. Elsheikh (Heriot-Watt University, EH14 4AS, United Kingdom)

**Abstract**

Quantitative workflows utilizing real-time data to constrain ahead-of-bit uncertainty have the potential to improve geosteering significantly. Fast updates based on real-time data are essential when drilling in complex reservoirs with high uncertainties in pre-drill models. However, practical assimilation of real-time data requires effective geological modelling and mathematically robust parameterization. We propose a generative adversarial deep neural network (GAN), trained to reproduce geologically consistent 2D sections of fluvial successions. Offline training produces a fast GAN-based approximation of complex geology parameterized as a 60-dimensional model vector with standard Gaussian distribution of each component. Probabilistic forecasts are generated using an ensemble of equiprobable model vector realizations. A forward-modelling sequence, including a GAN, converts the initial (prior) ensemble of realizations into EM log predictions. An ensemble smoother minimizes statistical misfits between predictions and real-time data, yielding an update of model vectors and reduced uncertainty around the well. Updates can be then translated to probabilistic predictions of facies and resistivities. The present paper demonstrates a workflow for geosteering in an outcrop-based, synthetic fluvial succession. In our example, the method reduces uncertainty and correctly predicts most major geological features up to 500 meters ahead of drill-bit.

**Introduction**

Sub-surface uncertainty is typically handled using multi-realization geo-models covering the expected range of spatial configurations and property distributions. Ideally, these uncertainties should be incorporated into geosteering workflows (Hermanrud et al., 2019). Such workflows require interaction between the pre-drill model and new information gained while drilling, which calls for methods that enable fast model updates and real-time forecasting based on continuously transmitted well data. The state-of-the-art workflows being implemented in the field can currently predict formation boundary a hundred meters ahead of bit (Antonsen et al. 2018).

The ensemble Kalman filters (EnKF) and smoothers (ES) and their iterative versions are efficient methods to sequentially update existing multi-realization geo-models with new real-time information, reduce the uncertainty, and create probabilistic forecasts with a decision recommendation (Chen et al. 2015, Alyaev et al. 2019). However, they require a conforming parametrization of the geo-model realizations for the computation of stochastic gradients. Thus, in the presence of discrete geological features, such as faults, additional ad-hoc procedures are needed to represent and update the relevant parameters (Chen et al., 2014), rendering the methodology slower and case-dependent. In practice, those geologically complex targets with elevated levels of uncertainty are likely to experience a greater need for updates during drilling workflows (Hermanrud et al., 2019), emphasizing the need for new automated workflows.

The key to unlocking the EnKF's / ES's potential for geosteering is creating mathematically well-behaving parametrizations for complex structural and facies models. Unsupervised learning for subsurface prediction and modelling has been gaining popularity across several applications, including nuclear-magnetic-resonance-log interpretation (Li and Misra 2017) and conditional facies modelling (Chan and Elsheikh 2019) and seismic analysis (Manouchehri et al. 2020). These methods are specifically advantageous in real-time applications where prediction/evaluation performance is critical. Fossum et al. (2021) proposed using a generative adversarial network (GAN) to produce complex geological realizations in real-time to aid geosteering. The authors verified that the chosen iterative ES reduces the uncertainty in the GAN-based multi-realization model. Moreover, the iterative ES results were compared to results from a Monte-Carlo method.

In this paper, we demonstrate forecasting of geological uncertainty using the GAN on a synthetic low net-gross fluvial reservoir, emphasizing the workflow's geological and operational aspects. We consider a geosteering operation across a complex 3D channel architecture. We first describe the methodology for offline training of the GAN, which ensures that all generated realizations are in-line with the selected geological setting and creates a well-behaved low-dimensional characterization for local geo-models. We then describe a workflow which uses the GAN in the modelling sequence and enables the ES Bayesian update. After that, we describe a synthetic example and show how the methodology is used to reduce uncertainties and predict geological features ahead of drilling. Finally, concluding remarks and suggestions for further work are outlined.

**Method**

In this section, we describe the main components of the proposed workflow. First, we describe the GAN trained to generate realizations of the subsurface. Second, we discuss the dataset which was used for training. Finally, we show how the GAN model can be used within a Bayesian workflow.

*GAN: structure and training*

In our workflow, we represent the local geo-model relevant for geosteering as an ensemble of equiprobable realizations. Each realization is parametrized by a vector of 60 parameters (model vector), for which a GAN learns the geological meaning during an offline training phase. Notably, the GAN model-vector inputs are parameterized as standard Gaussian, ensuring the mathematical robustness of the inversion.

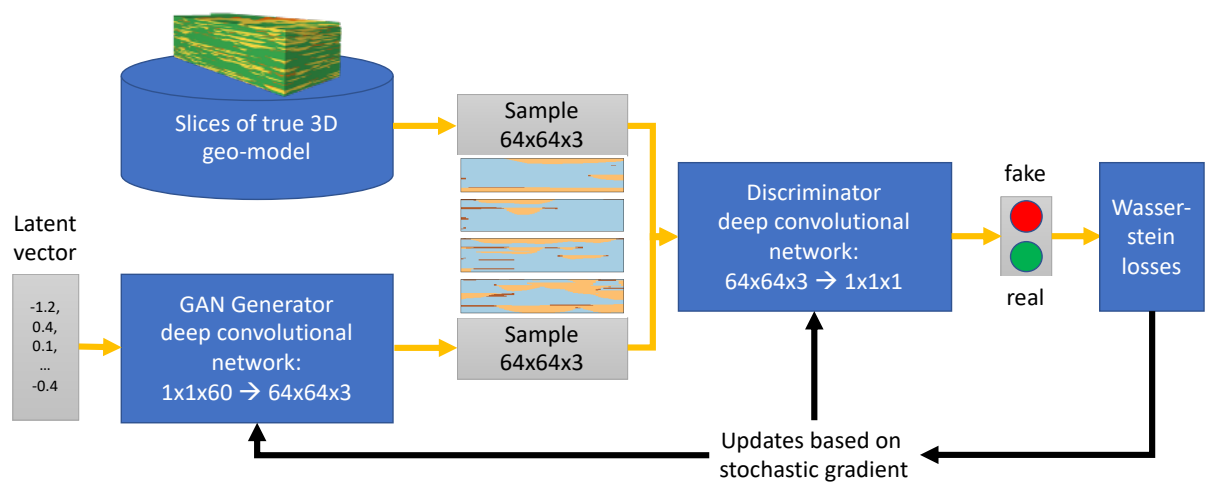

*Figure 1 Training of the GAN for generations of facies realizations in 2D.*

The GAN consists of two parts: the primary network, the generator, and the supplementary network, the discriminator, see **Figure 1**. The role of the generator is to convert a (standard Gaussian) model vector to a synthetic (aka. fake) data, in our case, a local geo-model, which looks like the actual training data. The role of the discriminator is to distinguish correctly between the fake and the real data, where real data is any geo-model from the training dataset. The two networks are trained simultaneously by playing a min-max game, in which each of them tries to reach its objective (Goodfellow et al. 2014). This study uses deep convolutional network architectures (DCGANs) for both the generator and the discriminator (Radford et al. 2016). For the loss function during training, we use the Wasserstein loss function proposed for GAN by Arjovsky et al. (2017).

*Training dataset*

The training dataset in the present study consists of 2D sections sampled from a 3D geo-model. The geo-model shown in **Figure 2** was constructed using a standard reservoir modelling tool and conforms to several predefined requirements intended to provide a realistic case while analyzing test results more transparent. It is based on empirical data, with a moderate to high complexity involving a limited number of architectural elements with contrasting properties.

The geo-model measures 4000m x 1000m x 200m (XYZ) and utilizes a local coordinates system. The vertical extent of the model is constrained by a flat top and bottom surface positioned at 1800m and 2000m, respectively. Grid resolution was set at 10m x 10m x 0.5m (XYZ), yielding 16 million grid cells. The facies set-up in the geo-model (**Table 1**) represents a synthetic low net/gross fluvial system, which provides complex 3D architectures with contrasting petrophysical properties. The system does not include faults or fractures.

Specification of facies geometries and dimensions are derived from a well-documented outcrop of the lower Williams Fork Formation of the Cretaceous Mesa Verde Group at Coal Canyon in Colorado (Pranter and Sommer, 2011; Pranter et al., 2014; Trampush et al., 2017). The outcrop at Coal Canyon exhibits a 1500 m wide and 200 m thick succession representing the fluvial coastal plain of a foreland basin with warm/temperate climatic conditions. Although the geo-model uses key facies parameters from the site, it is not intended as a detailed rendering of the outcrop itself. Our model is a simplification based on the works of Ellison (2004); Cole and Cumella (2005); Panjaithan (2006); Pranter et al. (2009), and Pranter et al. (2014). Our model contains only three facies: background/shale, channels, and crevasse splays. Our facies model set-up defines only single-storied channels, for which the probability distribution follows "narrow channel bodies" (Pranter et al. 2009, Pranter et al. 2014). Multi-storied channels are rendered in the model by stacking, see **Figure 2**. The flow direction of the channel system was set to 45+10 degrees, and their spatial distribution is not conditioned by any trends. Map-view geometry of crevasse splays is challenging to derive from outcrop observations. In the model set-up, the volumetric proportion of crevasse splays was used as a guideline. The remaining parameter settings were adjusted heuristically. Parameters used for the facies model set-up are summarized in **Table 1**.

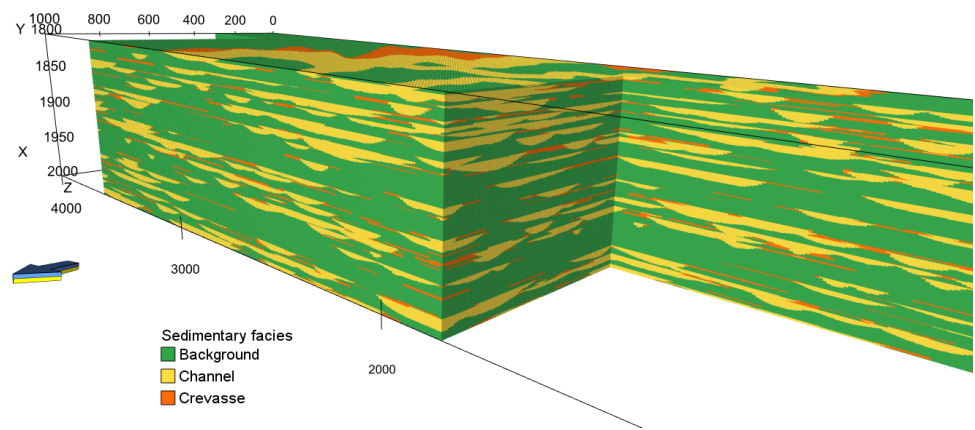

*Figure 2 The synthetic geo-model used for training of GAN in the present study, which is based on the parameters of the lower Williams Fork Formation of the Cretaceous Mesa Verde Group at Coal Canyon in Colorado.*

| Channels | | | Crevasse splays | | |
|---|---|---|---|---|---|
| **Parameter** | **Setting** | | **Parameter** | **Setting** | |
| Cahnnel form | rigid parabolic | | Crevasse geometry | isolated, no continuous belts | |
| Channel positioning | no trends, no repulsion | | Crevasse vertical position | towards top of channel | |
| | Value/mean | std | | Value/mean | std |
| Volume fraction | 0.3 | 0.05 | Volume fraction (relative to channel) | 0.3 | 0.05 |
| Thickness, m | 4.2 | 1.5 | Thickness (relative to channel) | 0.3 | 0.1 |
| Width, m (min=20, max=500) | 155.0 | 50.0 | Channel margin coverage | 0.5 | |
| Width/thickness relation | 36.0 | | Relative lobe width normal to channel | 0.7 | |
| Amplitude, m | 400.0 | 50.0 | Absolute lobe length parallel to channel, m | 200.0 | |
| Sinuosity | 1.3 | | Min. # of crev. belts per channel | 2 | |
| Azimuth, deg | 45.0 | 10.0 | Max. # of crev. belts per channel | 6 | |

*Table 1 Key parameter settings used in the facies model set-up.*

For the training, 2D XZ-sections measuring 64 x 64 cells are sampled from the 3D geo-model to create realizations of size 64 x 64 (x3): 32 m in depth and 640 m in length, with the third dimension representing the probability that the cell contains facies with the corresponding index. The selected dimensions of a realization are relevant to provide geosteering support but can be modified as needed for real operations.

*Bayesian update of the probabilistic model with EnRML*

EnKF / ESs use a discrete formulation of the Bayes theory to estimate the posterior distribution of the model parameters by conditioning the set prior realizations to the gathered measurements. To do so, we need a forward model that links the model parameters to the measurements.

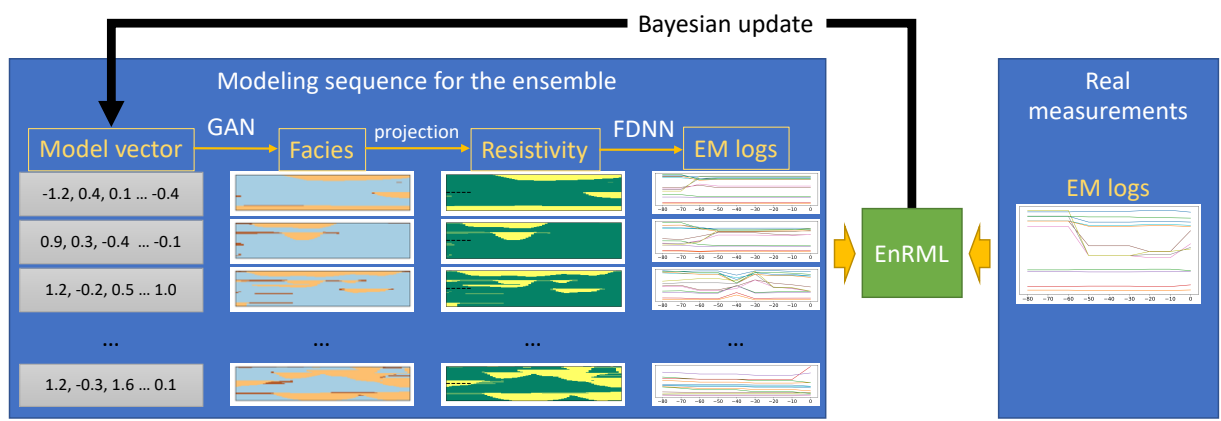

*Figure 3 The proposed EnRML workflow with GAN.*

For every realization in the ensemble, we use a three-step modelling sequence to convert the model vector to the EM-log forecasts, see

**Figure 3**. First, the GAN converts the model vector to a 2D geo-model. Second, the geo-model is transformed into a series of 1D resistivity projections at measurement locations. Third, a forward deep neural network (FDNN) converts the 1D inputs to synthetic EM logs: a full suite of extra-deep EM log traces transmitted in real-time during drilling (Alyaev et al., 2021). The extra-deep EM logs include 22 measurements with a sensitivity of up to 20 meters from the tool (Larsen et al. 2018). For one realization, the entire modelling sequence takes around 0.12s to model nine logging positions along the well path on a 3.3 GHz (4.5 GHz turbo) core.

In this paper, we use an iterative ES, namely, the approximate Levenberg-Marquardt ensemble randomized maximum likelihood method (EnRML) introduced by Chen and Oliver (2013), to condition the prior model ensemble to the real-time measurements. At every iteration, the method minimizes an objective function, representing a regularised statistical misfit between the expected and the actual measurements (Fossum et al., 2021). As a result, the ensemble of model vectors is updated to better match observations, see

**Figure 3**. The EnRML regularization is specifically important when working with highly non-linear forward models (Chen and Oliver 2013), and our tests indicate that the described modelling sequence is highly non-linear. The iteration process is terminated when the statistical misfit stops decreasing or after ten iterations to prevent over-fitting.

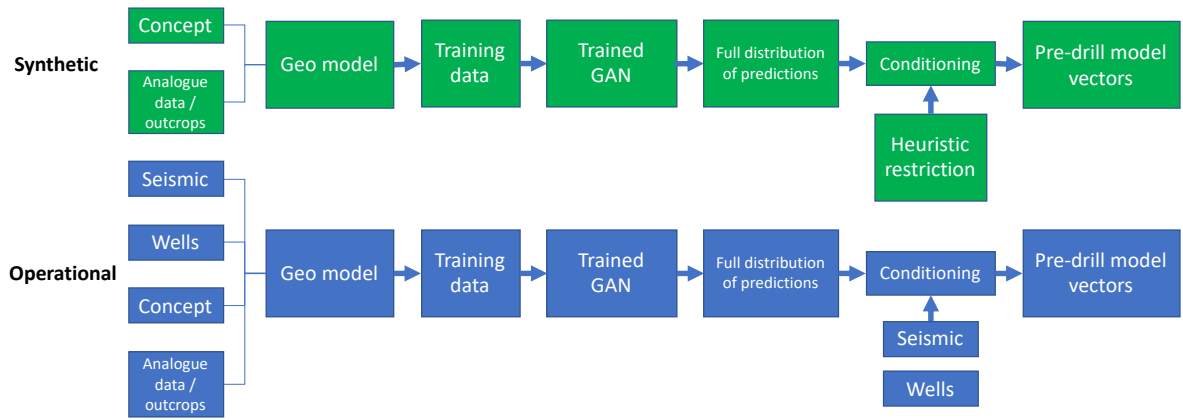

*Figure 4 The preparation of the pre-drill model vectors for the synthetic case and potential operational use.*

**Example**

To illustrate the workflow, we set up a synthetic example of geosteering in the fluvial succession rendered in the geo-model described above. We select a representative synthetic truth, which contains several channel groups surrounded by crevasse splays, see **Figure 5**. We make the simplifying assumption that the resistivity for each facies is known and constant, and the reservoir is oil-filled: background/ shale, 4.0 ohm m; oil-saturated channel sand, 171.0 ohm m; partially saturated crevasse splay sand, 55.0 ohm m. We use the same resistivity values and the same forward deep neural network model (Alyaev et al., 2021) to generate the true and the expected measurements from the ensemble.

A probabilistic pre-drill 2D geo-model with 500 realizations is constructed by randomly selecting 60-parametric model vectors from heuristically restricted prior. The heuristic prior has a Gaussian distribution with the standard deviation of 1.0 and the mean, which had 24 components set to 0.0 and the rest to 25% of the corresponding components of the synthetic truth vector. In practice, a pre-drill geo-model will be conditioned on information from surrounding wells and/or seismic data, see **Figure 4**. The mean of the resulting probabilistic model is not localized or even structurally like the synthetic truth, see **Figure 6**. At the same time, GAN guarantees that all individual realizations conform to geologic "reality".

In this study, we simulate drilling a horizontal well in the middle of the 2D geo-model. The measurements are acquired for the first 90 meters (nine cells with one measurement per cell) and are assimilated together. The modelled extra-deep tool is sensitive to up-to first three contrasts/boundaries up and down from each measurement location, see the arrows in **Figure 5**.

After ten iterations of EnRML updates, we arrive at the posterior probabilistic resistivity model, see **Figure 7**. Looking at the standard deviation, we see that the uncertainty around the well path is significantly reduced, about six meters up and ten meters down from the well. This is in line with the direct sensitivity of the extra-deep EM tool. Moreover, using the geological features learned by the GAN, the uncertainty reduction also apparently extends 150 meters ahead of the bit; see the reduction of std in **Figure 7**.

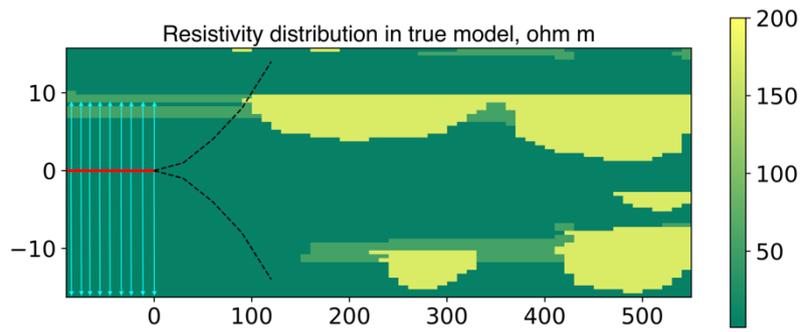

*Figure 5* Resistivity of the synthetic truth. The solid line indicates the position of the well, the arrows show the tool sensitivity, and the dashed lines show the possibility of steering.

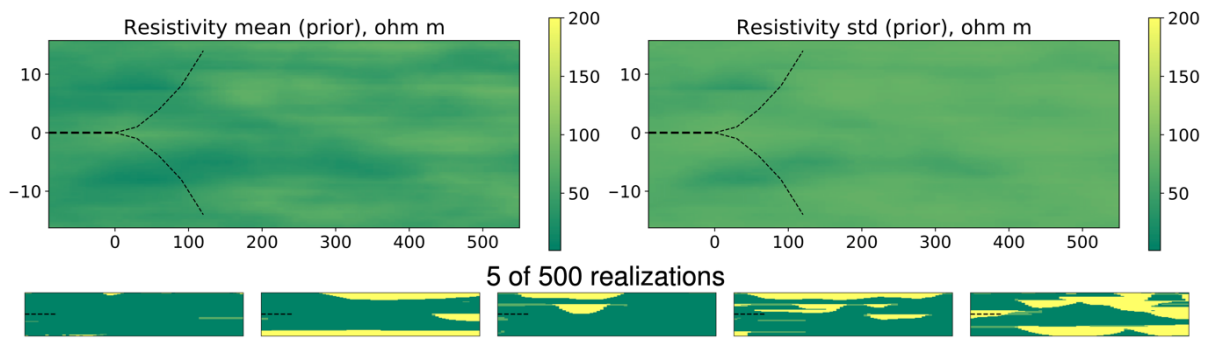

*Figure 6* The resistivity of the probabilistic prior/pre-drill model. The mean, the standard deviation, and five realizations are shown.

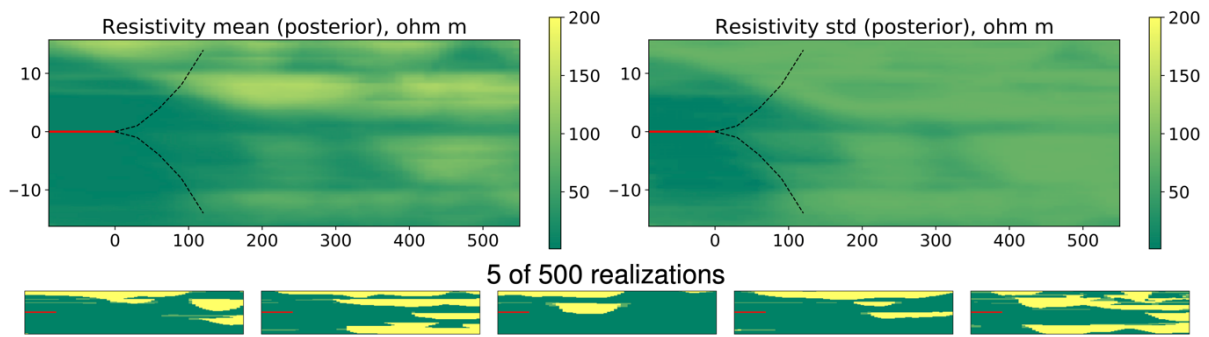

*Figure 7* The resistivity of the probabilistic posterior model (after update). The mean, the standard deviation, and five realizations are shown.

The five realizations shown in **Figure 7** all contain at least one thin crevasse splay above the well, to which deeper-sensing log traces have high sensitivity. Outside this area, the geo-model realizations preserve geological consistency but are pretty different from each other and the synthetic truth. Nevertheless, inspecting the mean of all 500 posterior realizations, one can observe a fuzzy outline of the geo-bodies from the synthetic geo-model, specifically the top group of channels surrounded by crevasse splays.

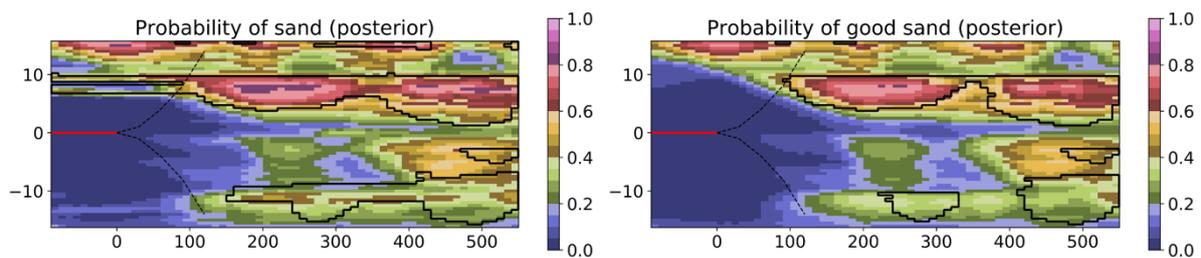

*Figure 8* Map of the posterior probability of target sands and the (good) channel sands. Outlines mark the facies' boundaries from the synthetic truth.

To further highlight the predictive capability of the workflow, we consider the probability of the target facies, see **Figure 8**. Based on the log data, background shale is identified from 6 to 15 meters around the well and predicted to extend with a high probability up to 200 meters ahead of the last measurement position. The two crevasse splays above the well, indicated by the measurements, are also identified. Probability values around 0.5 and one-cell deviation from the synthetic truth can be attributed to the uncertainty of the boundaries for those thin layers. Based on these extended crevasse splays, the workflow correctly forecasts the two connected large channels located to the right of the splays. Predictably, as the uncertainty grows away from the measurements (**Figure 7**, std), the quality of the sand body shape predictions becomes worse (**Figure 8**). Nevertheless, the vertical stacking of channels is correctly forecasted to the right of the geo-model, 500 meters ahead of the bit. With such limited data, the geometry and location of that geo-body are not predicted well. Similarly, the bottom succession of channels, connected by crevasse splays, is also qualitatively predicted despite a lack of data.

Other features are forecasted less precisely. The top-most channel group in the posterior is much more pronounced than only a few cells in the synthetic truth, see **Figure 8**. This can originate from the heuristic centring of the prior around zero, where sands at the top of the geo-model are very probable, see **Figure 6**. The not-so-probable channel stack in the middle of the geo-model (**Figure 8**) has a sufficiently low probability of influencing decision-making.

**Conclusions**

In this communication, we showed that the EnRML workflow equipped with GAN could reliably forecast complex fluvial succession ahead of drilling. Currently, the workflow outputs probabilistic predictions of facies and resistivities 500 meters ahead of the logging tool. As shown in the numerical experiment, the posterior model has reduced uncertainty and quantitatively predicts most of the geological features for the entire model.

The main computational burden of complex geological modelling is taken during the offline training of GAN. The online modelling sequence, as well as the EnRML inversion, is suitable for real-time geosteering even with larger ensemble sizes. The GAN-based probabilistic workflow can be naturally extended to other geological settings. We expect the GAN to handle other structural and sedimentary uncertainties with appropriate training.

The results of our method for synthetic cases show good probabilistic prediction of both discrete and continuous features at distances exceeding the operational state of the art (Antonsen et al. 2018). Ensuring forecasts of comparable quality under operational conditions will require moving from a heuristic prior to model vectors conditioned on hard data available on the pre-drill stage (using methods from, e.g., Chan and Elsheikh, 2019). To capitalize on the more complex predictions, global well-placement optimization algorithms, such as in Alyaev et al. (2019), should be integrated into the suggested workflow.

**Acknowledgements**

The researchers at NORCE are supported by the Centre for Research-based Innovation DigiWells: Digital Well Centre for Value Creation, Competitiveness and Minimum Environmental Footprint (NFR SFI project no. 309589, DigiWells.no). The centre is a cooperation of NORCE, UiS, NTNU, and UiB.

It is financed by the Research Council of Norway, Aker BP, ConocoPhillips, Equinor, Lundin, Total, and Wintershall Dea.

Part of the work was performed within the project' Geosteering for IOR' (NFR-Petromaks2 project no. 268122), which is funded by the Research Council of Norway, Aker BP, Equinor, Vår Energi and Baker Hughes Norway.

We thank Emerson Roxar for providing an academic license for RMS 11.1 used for the geo-modelling in this study.